\documentclass[%
reprint,
 amsmath,amssymb,
 aps,
 prl,
floatfix,
]{revtex4-1}
\usepackage{graphicx}% Include figure files
\usepackage{dcolumn}% Align table columns on decimal point
\usepackage{bm}% bold math
\usepackage{color}
\usepackage{tikz}
\usetikzlibrary{shapes}
\usepackage{amssymb}
\usepackage{floatrow}
\usepackage[caption=false]{subfig}
\begin{document}
\newcommand{\markerone}{\raisebox{0.6pt}{\tikz{\node[draw=red,scale=0.4,circle,fill=red!20!red](){};}}}
\newcommand{\markertwo}{\raisebox{0.8pt}{\tikz{\node[draw=blue,scale=0.3,regular polygon, regular polygon sides=4,fill=blue!20!blue](){};}}}
\newcommand{\markerthree}{\raisebox{0.8pt}{\tikz{\node[draw=blue,scale=0.5,regular polygon, regular polygon sides=4](){};}}}
%\preprint{APS/123-QED}

\title{Transition to turbulence in particle laden flows}% Force line breaks with \\
%\thanks{A footnote to the article title}%

\author{Nishchal Agrawal}
 \affiliation{ IST Austria, 3400 Klosterneuburg, Austria}
% \altaffiliation[Also at ]{Physics Department, XYZ University.}%Lines break automatically or can be forced with \\
\author{George H. Choueiri}%
 \affiliation{ IST Austria, 3400 Klosterneuburg, Austria}
 
\author{Bj\"orn Hof}
 \email{bhof@ist.ac.at}
\affiliation{ IST Austria, 3400 Klosterneuburg, Austria}

%\collaboration{MUSO Collaboration}%\noaffiliation

%\author{Charlie Author}
% \homepage{http://www.Second.institution.edu/~Charlie.Author}
%\affiliation{
% Second institution and/or address\\
% This line break forced% with \\
%}%
%\affiliation{
% Third institution, the second for Charlie Author
%}%
%\author{Delta Author}
%\affiliation{%
% Authors' institution and/or address\\
% This line break forced with \textbackslash\textbackslash
%}%

%\collaboration{CLEO Collaboration}%\noaffiliation

\date{\today}% It is always \today, today,
             %  but any date may be explicitly specified

\begin{abstract}
Suspended particles can alter the properties of fluids and in particular also affect the transition from laminar to turbulent flow. In the present experimental study, we investigate the impact of neutrally buoyant, spherical inertial particles on transition to turbulence in a pipe flow. At low particle concentrations, like in single phase Newtonian fluids, turbulence only sets in when triggered by sufficiently large perturbations and, as characteristic for this transition localized turbulent regions (puffs) co-exist with laminar flow. In agreement with earlier studies this transition point initially moves to lower Reynolds number (Re) as the particle concentration increases. At higher concentrations however the nature of the transition qualitatively changes: Laminar flow gives way to a globally fluctuating state following a continuous, non-hysteretic transition. A further increase in Re results in a secondary instability where localized puff-like structures arise on top of the uniformly fluctuating background flow. At even higher concentration only the uniformly fluctuating flow is found and signatures of Newtonian type turbulence are no longer observed.

%and the critical Reynolds number above which turbulence can be sustained decreases with particle concentration.  
%\begin{description}
%\item[Usage]
%Secondary publications and information retrieval purposes.
%\item[PACS numbers]
%May be entered using the \verb+\pacs{#1}+ command.
%\item[Structure]
%You may use the \texttt{description} environment to structure your %abstract;
%use the optional argument of the \verb+\item+ command to give the %category of each item. 
%\end{description}
\end{abstract}

\pacs{Valid PACS appear here}% PACS, the Physics and Astronomy
                             % Classification Scheme.
%\keywords{Suggested keywords}%Use showkeys class option if keyword
                              %display desired
\maketitle

%\tableofcontents

%\section{\label{sec:level1}First-level heading}

Particle laden flows are ubiquitous in nature and applications, including slurry flows, sediment transport, blood flow and pollutant dispersion in atmospheric flows. Particle-fluid interactions strongly affect the dynamics especially when the particles are sufficiently large i.e. larger than the smallest scales of the flow, and the inertial effects become important \cite{guazzelli2011physical}. 
In particular this also influences the transition from laminar to turbulent flow and the nature of turbulence. However, at present, the phenomenon of laminar-turbulent transition for particle laden flows is poorly understood, especially compared to a single phase Newtonian fluid.  

In case of a Newtonian, single phase fluid in a pipe, the laminar flow is linearly stable for all Reynolds numbers ($Re = \rho UD/\mu$) \cite{drazin2004hydrodynamic,meseguer2003linearized}, yet turbulence can be triggered if Re is sufficiently large. Finite amplitude perturbations lead to an abrupt onset of turbulence \cite{darbyshire1995transition,hof2003scaling}. At the lowest Reynolds numbers where turbulence is first encountered it only occurs in localized patches, so called puffs which are spatially separated by laminar flow \cite{eckhardt2007turbulence}. The coexistence of laminar and turbulent states i.e. spatio-temporal intermittency, and the dependence of the critical Reynolds number for transition on the strength of perturbations (and hence hysteresis), are characteristic for transition in Newtonian, single phase flow.  

Adding particles to the fluid can significantly alter this scenario due to particle-fluid and particle-particle interactions \cite{mueller2009rheology}. Matas \textit{et al.} \cite{matas2003transition} investigated the effect of neutrally buoyant inertial particles on the laminar-turbulence transition in a pipe flow. They presented the critical Reynolds number $Re_c$ at which puffs were first detected in the flow, for varying particle concentrations and sizes and showed that for sufficiently large particles $Re_c$ varied non-monotonically with particle concentration. With the initial increase in particle concentration, transition was triggered at a progressively lower $Re_c$. However, upon further increase, unexpectedly, the trend reverses and $Re_c$ starts to increase. More recently, Yu \textit{et al.} \cite{yu2013numerical} reported similar non-monotonic behaviour in a numerical study. They also noted that it was difficult to rigorously judge whether the flow is laminar or turbulent, as velocity fluctuations increased smoothly with $Re$. In another numerical study, for neutrally buoyant spherical particles in a channel flow, Lashgari \textit{et al.} \cite{lashgari2014laminar} showed the existence of three different flow regimes: `laminar-like' regime that occurs at low-concentrations and low Re, `turbulent-like' regime at low concentrations and high Re, and `shear-thickening' regime at high concentrations and high Re. For the latter regime, the wall friction increased with Re due to particle induced stresses, while turbulent transport was weakly affected. Due to this, they speculated that, at high enough particle concentration the transition to turbulence might not only be delayed, as reported by Matas \textit{et al.} \cite{matas2003transition} but could be completely suppressed. Both Yu \textit{et al.} \cite{yu2013numerical} and Lashgari \textit{et al.} \cite{lashgari2014laminar} noted that the velocity fluctuations increased smoothly with Re at high particle concentration. Newtonian type turbulence on the other hand is accompanied  by a sharp increase in velocity and pressure fluctuations.

Several studies addressing dilute polymeric flow have also reported smoothly increasing velocity and pressure fluctuations at higher polymer concentration \cite{samanta2013elasto,choueiri2018exceeding}, here, transition to turbulence occurred without hysteresis or intermittency. They also noted that the ordinary Newtonian turbulence was suppressed and replaced by a different kind of disordered motion called elasto-inertial turbulence. Hence, it is possible that also in the case of particles smoothly increasing velocity and pressure fluctuations, observed in numerics and experiments \cite{matas2003transition,lashgari2014laminar,yu2013numerical,wen2017experimental}, indicate that the nature of transition and perhaps the turbulent state itself are changed by the presence of particles. 

In this paper, we focus our attention on experimentally exploring the mechanism behind laminar-turbulent transition for suspensions of neutrally buoyant spherical particles in a pipe flow. We show that, at high particle concentrations, indeed a different type of instability exists which occurs without any intermittency or hysteresis. The instability threshold decreases monotonically with particle concentration and can appear at much lower Reynolds numbers as compared to transition to Newtonian turbulence in single phase flows.  

The experimental set-up consists of a straight, horizontal glass tube of circular cross section with diameter $D = 4 mm$ and total length of $500 D$. Measurements were performed $300 D$ downstream from the inlet. Here, the pressure was measured over a length of $L = 120 D$ using a differential pressure sensor. Just downstream of that, another differential sensor is used over a length of $5 D$ to measure fluctuating quantities. Two different types of perturbations are used. First, to generate turbulent puffs, a $20 ms$ impulse injection is used $180 D$ downstream of the inlet; second, a continuous perturbation is used, in the form of a pin, $0.9 mm$ in diameter, located $15 D$ downstream of the inlet. The fluid used was a $21.65\%$ glycerin-water solution matching the density $\rho = 1.051 \pm 0.01 g.cm^{-1}$ of the suspended polystyrene spheres of diameter $d = 0.200 \pm 0.15 mm$. Consequently the pipe to particle diameter ratio  $D/d \approx 20 $. The suspension was driven by a piston set-up to ensure a constant volumetric flow rate.     

\begin{figure}[!htp]
	\centering
	\includegraphics[width= \textwidth]{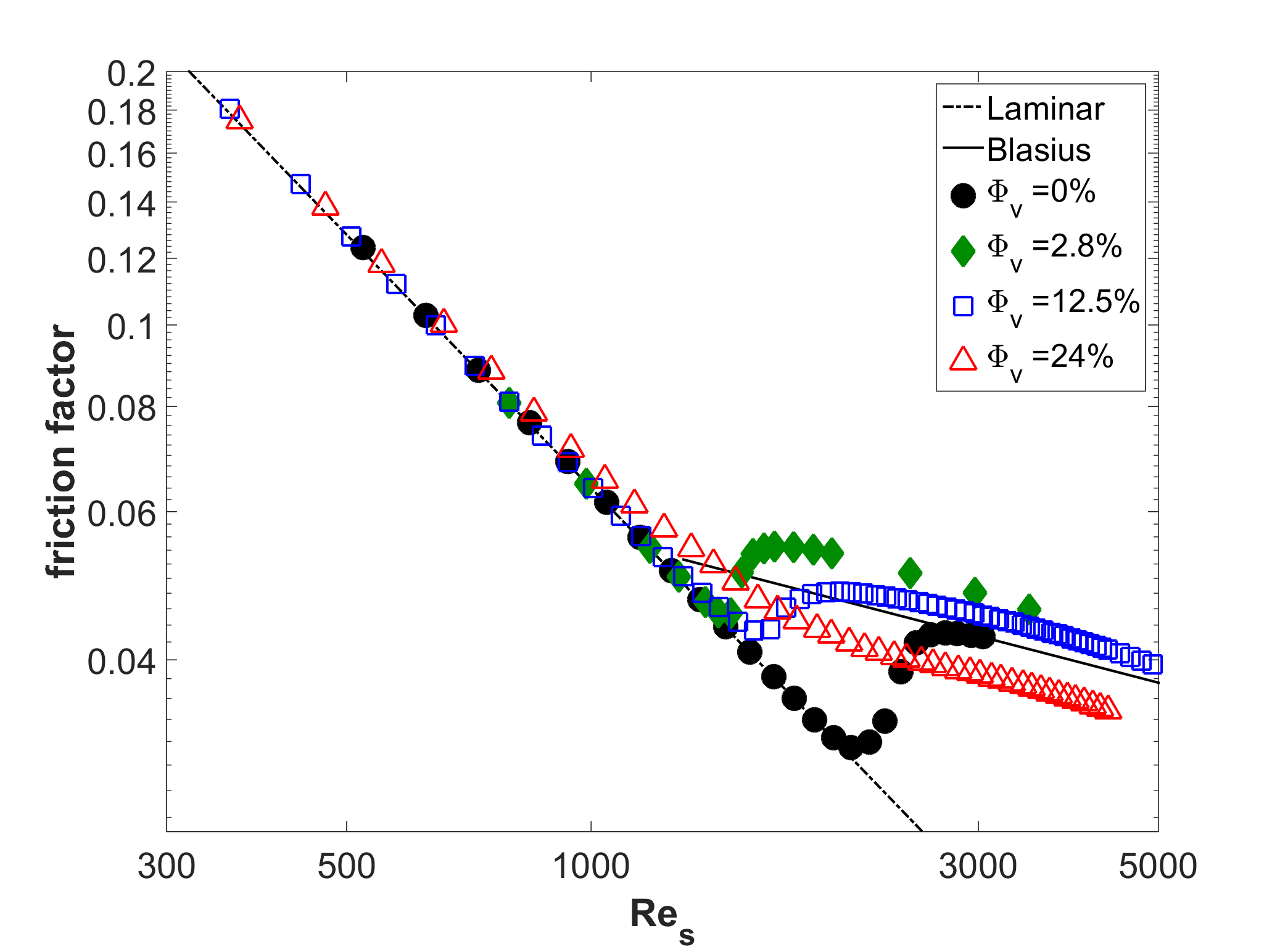}
	\caption{Friction factor as a function of suspension Reynolds number. Experiments were done in the presence of the continuous perturbation.}
	\label{FFWhole}
\end{figure}    

For all the plots shown in the paper we have used suspension Reynolds number $Re_s = \rho UD/\mu_{eff}$, where $U$ is bulk velocity and $\mu_{eff}$ is the effective dynamic viscosity of the suspension which is determined for each concentration $\Phi_v$ by collapsing measured pressure drop values onto the Hagen-Poiseuille curve when the flow is in the laminar state. The viscosity of suspensions of spherical particles is known to depend on $\Phi_v$ but for $\Phi_v \lesssim 25 \%$ the behaviour is still approximately Newtonian \cite{mueller2009rheology}. This is confirmed in Fig.~\ref{FFWhole} by the excellent collapse of the friction factor $f=2\Delta PD/(L\rho U^2)$ on the laminar line with a constant $\mu_{eff}$ used for each concentration. $\Delta P$ is pressure drop across length $L$.

\begin{figure}[!htp]
	\centering
	\includegraphics[width= \textwidth]{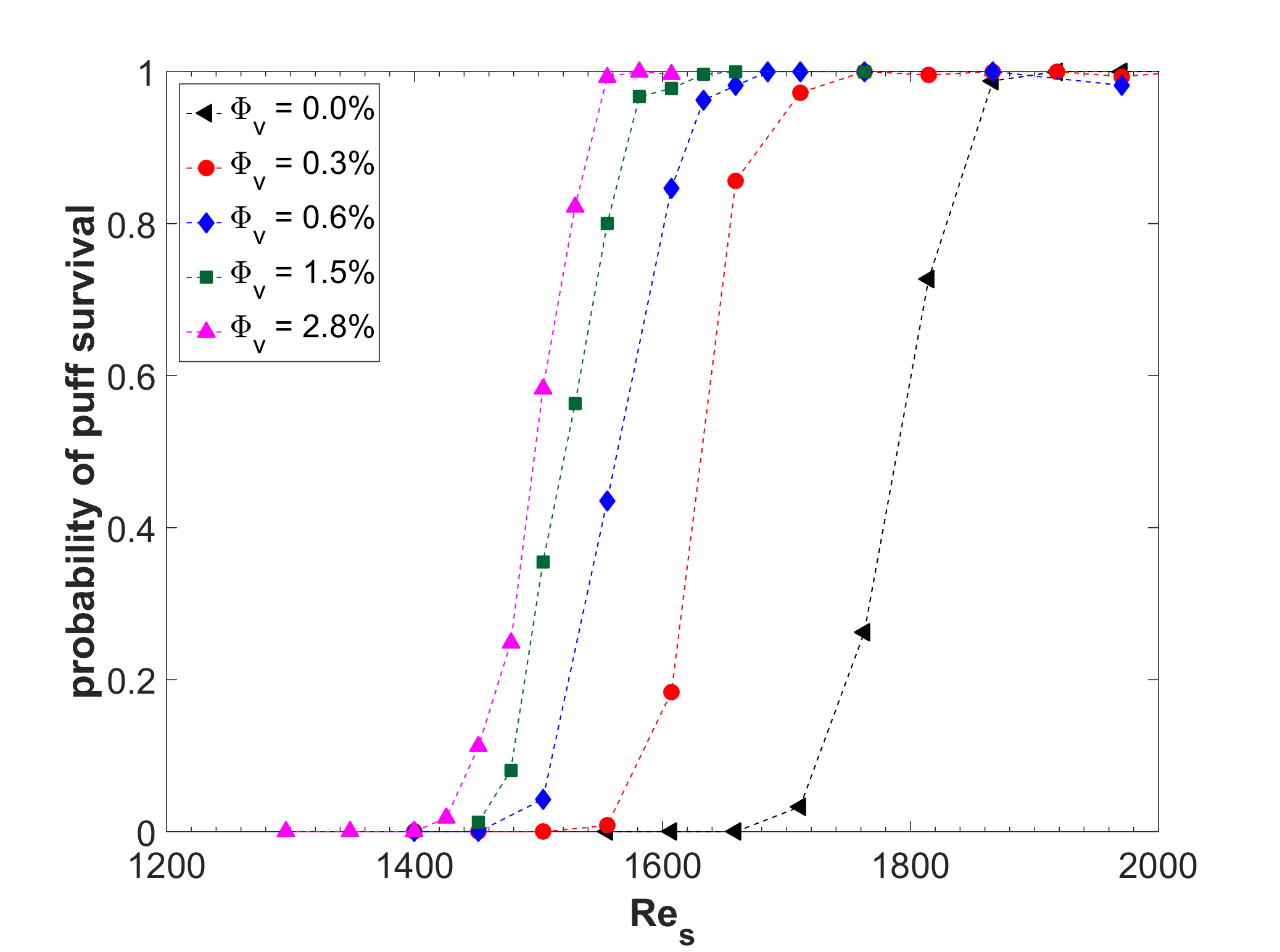}
	\caption{Survival probability of puff as a function of Reynolds number. Impulse injection was used to generate puffs. Measurement were taken approximately $263 D$ downstream of the perturbation point.}
	\label{Lifetime}
\end{figure}	

\begin{figure}[!htp]
	\centering
	\includegraphics[width= \textwidth]{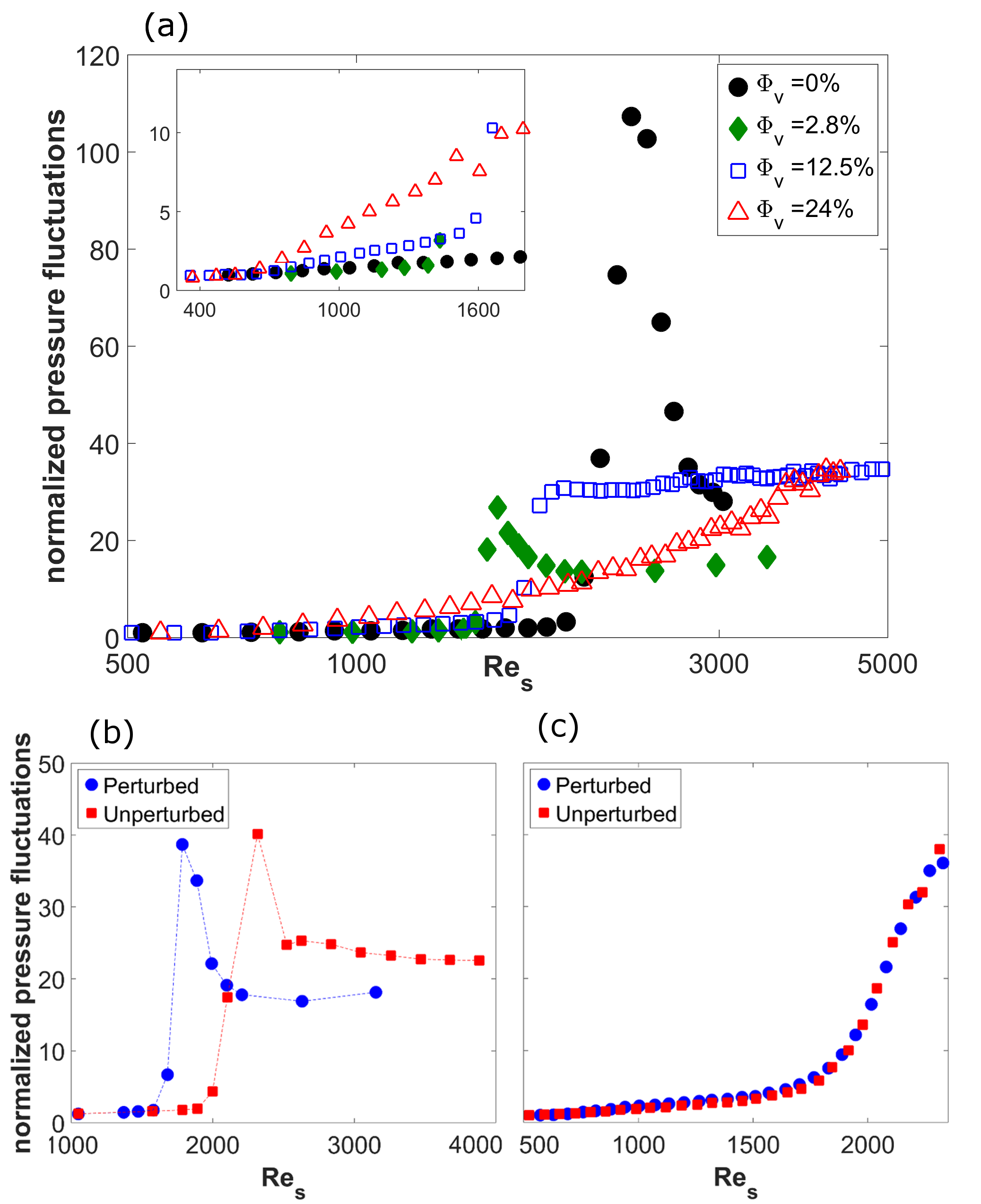}
	\caption{Normalized pressure fluctuations as a function of suspension Reynolds number, (a) for different concentrations in presence of the perturbation, (b) $\Phi_v = 0.6\%$ showing hysteresis, and (c) $\Phi_v = 16\%$ showing no hysteresis. }
	\label{Prms2}
\end{figure}

%\floatsetup[figure]{style=plain,subcapbesideposition=top}
%\begin{figure}[!htp]
%	\centering
%	\sidesubfloat[!htp]{
 %   \includegraphics[width=0.85\textwidth]{Figures/histeresisSub}
%    \label{HysSub}
%    }
%    \newline
%    \sidesubfloat[!htp]{
%    \includegraphics[width=	0.85\textwidth]{Figures/histeresisSuper}
%    \label{HysSup}
%    }
%    \caption{Normalised pressure fluctuation as function of $Re_s$ $(a)$ for $\Phi_v = 0.6\%$ hysteresis is observed, $(b)$ for $\Phi_v = 16\%$ no hysteresis is observed. For perturbed case continuous perturbation is used. }
%\end{figure}
 
\floatsetup[figure]{style=plain,subcapbesideposition=top}
\begin{figure}[!htp]
	\centering
	\sidesubfloat[!htp]{
    \includegraphics[width=0.9\textwidth]{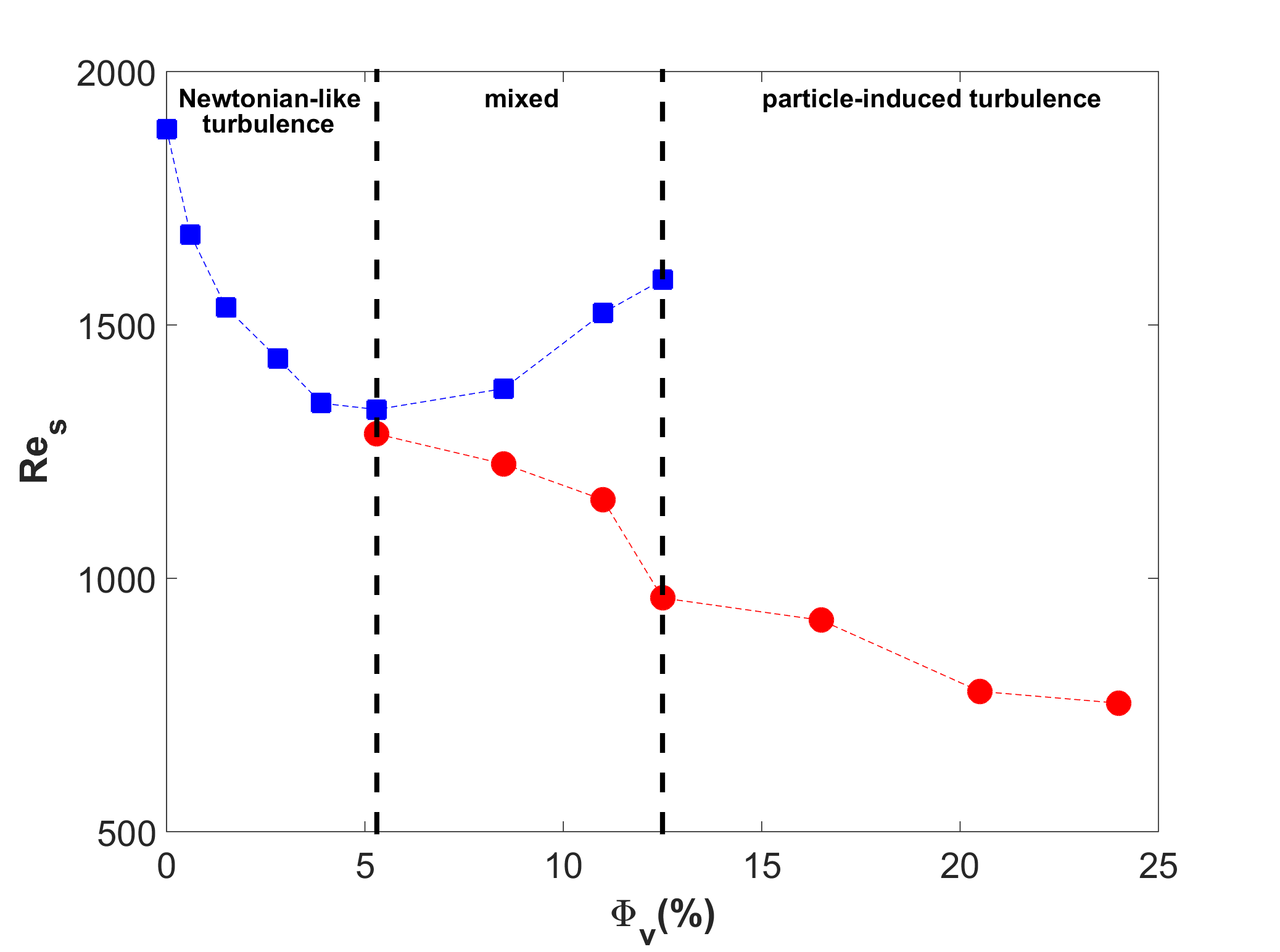}
    \label{Trans}
    }
    \newline
    \sidesubfloat[!htp]{
    \includegraphics[width=	0.9\textwidth]{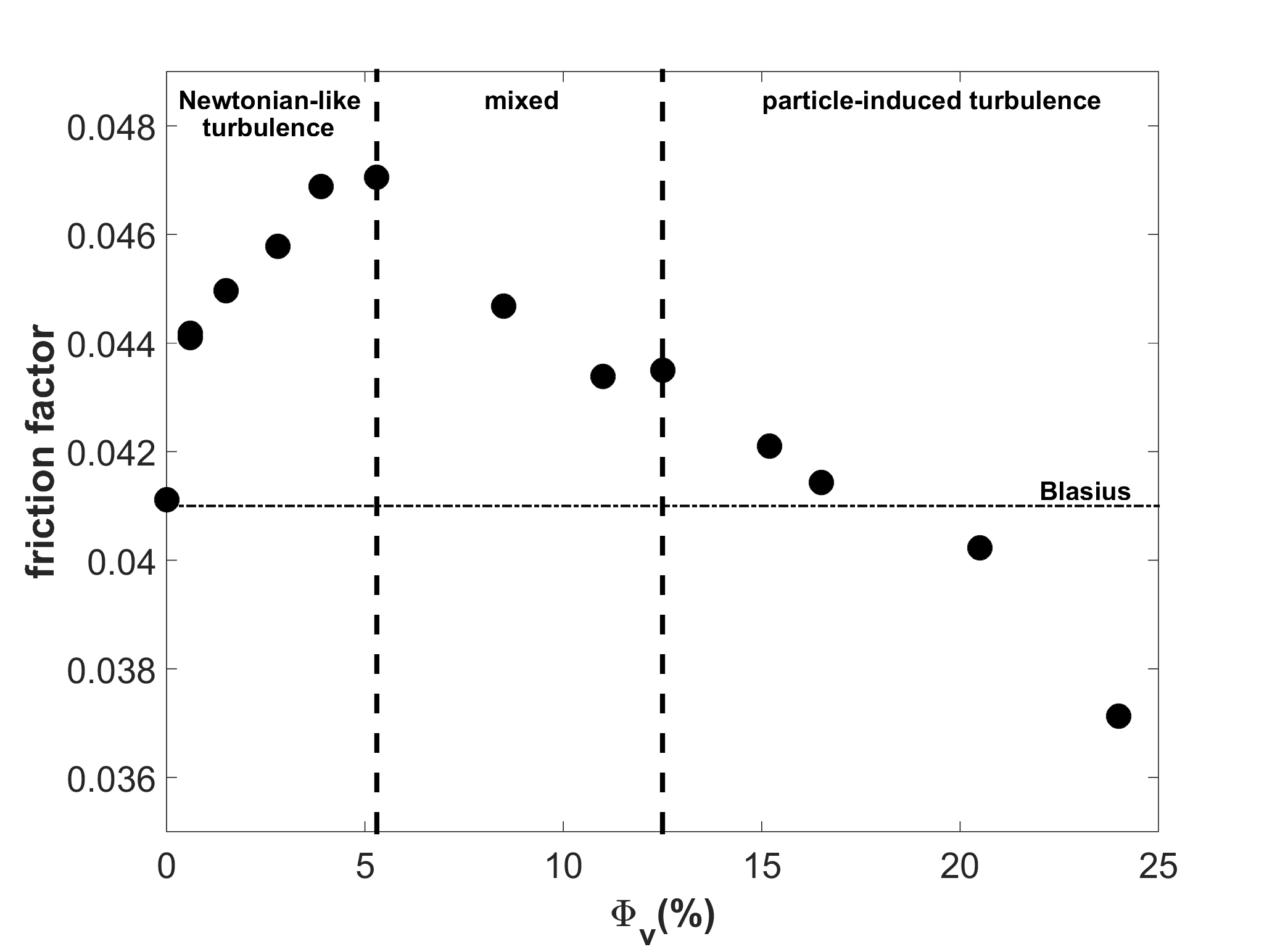}
    \label{FF}
    }
    \caption{(a) Transition scenario as a function of particle concentration: Newtonian-like (\protect\markertwo),  Particle-induced (\protect\markerone). (b) Friction factor at a fixed Reynolds number of $Re_s = 3500$ as a function of particle concentration. Experiments were done in the presence of the continuous perturbation.}
\end{figure}
    
%\begin{figure}[!htp]
%	\centering
%	\includegraphics[width= 8.3cm]{Figures/Transition}
%	\caption{Transition scenario as a function of particle concentration. Newtonian-like (\protect\markertwo),  Particle-induced (\protect\markerone)}
%	\label{Trans}
%\end{figure}

%\begin{figure}[!htp]
%	\centering
%	\includegraphics[width= 8.3cm]{Figures/FF3500}
%	\caption{Friction factor as a function of particle concentration for $Re_s = 3500$. Experiments were done in the presence of the continuous perturbation.}
%	\label{FF}
%\end{figure}    

% 
Puffs, characteristic for turbulence at low $Re$ in ordinary Newtonian fluids, have only a finite lifetime at any $Re$ \cite{hof2006finite} and so, given sufficient time, these structures can decay back to laminar, following a memoryless process \cite{faisst2004sensitive,peixinho2006decay}. Therefore, at a given $Re$ there is a distinct probability that a turbulent puff will survive beyond a certain time horizon. To quantify the effect of particles on transition we show the survival probability of the turbulent puffs in Fig.~\ref{Lifetime}. Like in the case of an ordinary Newtonian fluid, the survival probability increased with $Re_s$ for all concentrations following an S--shaped curve \cite{hof2008repeller}.
With increasing particle concentration, the curves shift to the left, and hence particles cause an earlier transition. For $4\% \lesssim \Phi_v \lesssim 12.5\%$, although the puff-like intermittent structures were observed in the flow, unexpectedly, we failed to generate puffs by perturbing the flow using impulse injections, irrespective of the perturbation strength. Therefore, it was not possible to collect survival probability for higher concentrations. To get a quantitative comparison at these higher concentrations we measured friction factors (Fig. \ref{FFWhole}) and pressure fluctuations (Fig. \ref{Prms2}).     

Normalized pressure fluctuations $p'/p'_o$ for different particle concentrations are plotted in Fig.~\ref{Prms2}a, where $p'$ is the standard deviation of pressure measured over the $5D$ distance, and $p'_o$ is the standard deviation due to background noise. As expected, the normalized pressure fluctuations are close to one when the flow is laminar. For $\Phi_v = 0$, fluctuations increase steeply at the onset of turbulence. Here the flow intermittently changes between laminar and turbulent regions which causes the high fluctuation levels.  As $Re_s$ is increased the turbulent fraction rises until the flow is fully turbulent. This behaviour is similar for concentrations up to $5\%$ although the fluctuation peak becomes less pronounced and moves to lower $Re_s$ (in line with the puff lifetime studies, shown in Fig.~\ref{Lifetime}). However for concentrations larger than $5\%$, weak but uniform fluctuations are observed which increase steadily with $Re_s$, atypical of Newtonian turbulence. For intermediate concentrations ($5\% \lesssim \Phi_v \lesssim 12.5\%$), signatures of localized puff-like structures are found as the $Re_s$ is further increased sufficiently far above the onset of the weakly fluctuating state. The Reynolds number for onset of puff in this regime increases with concentration. This shows that, the critical Re, where puffs first appear, varies non-monotonically with particle concentration, and the transition point first decreases and then increases again. This observation is in line with Matas \textit{et al.} \cite{matas2003transition} and Yu \textit{et al.} \cite{yu2013numerical}. For concentrations higher than $12.5\%$, no spatio-temporal intermittent puff-like structures could be found at any $Re_s$. Here, the continuous transition to the fluctuating state is found at even lower $Re_s$ and with increasing $Re_s$ fluctuation levels increase uniformly throughout space. These fluctuations can be observed for $Re_s$ as low as 800, which is far below the lowest Re where turbulence is first observed for Newtonian, single phase pipe flow.  

To probe the dependence of the transition point on perturbation levels and if it is hysteretic we compared measurements with and without the continuous perturbation. As can be seen in Fig.~\ref{Prms2}b transition is hysteretic at low concentrations and in the presence of the perturbation turbulence appears earlier than in the unperturbed case. In contrast, at high concentrations (see Fig.~\ref{Prms2}c), transition occurs at a specific Reynolds number value regardless of whether the fluid is perturbed or not. The transition in this case is continuous and pressure fluctuations feature neither an abrupt jump nor an initial overshoot (these being characteristic of spatio-temporal intermittency). The insensitivity of the transition to the finite amplitude perturbations and the smooth and continuous increase in fluctuations with increase in Re and uniformly fluctuating flow suggests that this type of transition (lower branch shown by (\protect\markerone) in Fig.~\ref{Trans}) may correspond to a linear-instability of the laminar base flow. 

Fig.~\ref{Trans} depicts transition thresholds for the two different types of instabilities encountered. The first branch i.e. `Newtonian-like', is caused as a result of a finite amplitude perturbation and varies non-monotonically with $\Phi_v$. The second branch (potentially corresponding to a linear instability), we denote as `particle-induced branch', is only detected for $\Phi_v \gtrsim 5\%$ and decreases monotonically with increasing concentration. However, overall, the transition threshold (either finite amplitude or linear instability) decreases monotonically with Re. 

Based on the existence of the two instabilities, we propose three different regimes. The `Newtonian-like turbulence' regime exist for $\Phi_v < 5\%$. The transition is abrupt, intermittent and extremely sensitive to perturbations. In this regime $Re_{cr}$ decreases with increasing concentration. For $5\% \lesssim \Phi_v \lesssim 12.5\%$, there exists a `mixed' regime where both branches exist. First, the turbulent-fluctuations appear globally and increase in intensity as Re is increased. On further increase in Re a secondary transition is eventually encountered and turbulent-puff like intermittent structures appear in the flow. However, no signs of hysteresis was detected. Unlike in `Newtonian-like turbulence' regime, here $Re_{cr}$ for the Newtonian-like branch increases with increase in $\Phi_v$. Interestingly, the trend of the Newtonian branch occurs at about the same concentration when the first signs of the particle-induced instability are observed in the flow and therefore could also be the cause of it. For $\Phi_v > 12.5\%$, we encounter the `particle-induced turbulence' regime where laminar flow gradually become turbulent with increasing $Re_s$. The flow is neither intermittent nor hysteretic and turbulent fluctuations can be seen for successively lower $Re_s$ as we increase the concentration.

To examine how the drag is modified by the presence of particles, we go back to Fig.~\ref{FFWhole}. For zero concentration, $f$ starts to increase at the onset and reaches Blasius when fully turbulent. Interestingly, for concentrations of $2.8\%$, $f$ for fully turbulent flow is higher than Blasius even though, as noted before in Fig.~\ref{Prms2}a, the peak value of pressure fluctuation as well as pressure fluctuations when the flow is fully turbulent is significantly lower. It is even more interesting to note that $f$ for fully turbulent flow has a non-monotonic dependence on concentration. To further elaborate this, $f$ is plotted in Fig.~\ref{FF} as a function of particle concentration for the $Re_s = 3500$, where the flow is fully turbulent for all concentrations. First, $f$ increases with $\Phi_v$, reaching a maximum for around $5\%$ and then the trend reverses with further increase in $\Phi_v$. Curiously, as in case of Newtonian-like branch in Fig.~\ref{Trans}, the trend reversal occurs around the same concentration when we first encounter signs of particle-induced turbulence. Furthermore, for $\Phi_v > 16\%$, $f$ falls below that of a single phase fluid having a viscosity equal to $\mu_{eff}$, implying drag reduction. Drag reduction as compared with viscosity matching Newtonian fluids is observed only in the particle-induced turbulence regime.              

%%%%%%%%% Conclusion %%%%%%%%%%%%%%%%

% Restatement and Implication
In summary, we have uncovered a continuous instability of the laminar base flow, previously unknown for particle laden flows. The critical Reynolds number for which the instability first appears decreases monotonically with increase in concentration. Furthermore, particle-induced turbulence, at sufficiently high concentrations, can lead to lower drag compared to ordinary turbulence. 

\end{document}